\documentclass[12pt]{article}
\usepackage{graphics}
\usepackage{amssymb}

\renewcommand{\theequation}{\arabic{section}.\arabic{equation}}

\newcommand{\ie}{{\em i.e.}}
\newcommand{\eg}{{\em e.g.}}
\newcommand{\QED}{\mbox{\rule[-1.5pt]{6pt}{10pt}}}
\newcommand{\lhs}{l.h.s. }
\newcommand{\rhs}{r.h.s. }
\newcommand{\im}{{\rm Im\,}}
\newcommand{\D}{{\rm d}}
\newcommand{\e}{{\rm e}}
\newcommand{\sgn}{{\rm sgn\,}}
\newcommand{\R}{\mathbb{R}}
\newcommand{\Z}{\mathbb{Z}}
\newcommand{\DD}{{\cal D}}
\newcommand{\EE}{{\cal E}}
\newcommand{\FF}{{\cal F}}
\newcommand{\OO}{{\cal O}}
\newcommand{\RR}{{\cal R}}
\newcommand{\SSS}{{\cal S}}
\newtheorem{claim}{Claim}[section]
\newtheorem{theorem}[claim]{Theorem}
\newtheorem{proposition}[claim]{Proposition}
\newtheorem{lemma}[claim]{Lemma}
\newtheorem{remark}[claim]{Remark}

\begin{document}

\title{Geometrically induced spectrum in curved leaky wires}
\author{P.~Exner$^{a,b}$ and T.~Ichinose$^{c}$}
\date{}
\maketitle
\begin{quote}
{\small \em a) Department of Theoretical Physics, Nuclear Physics
Institute, \\ \phantom{e)x}Academy of Sciences, 25068 \v Re\v z,
Czech Republic \\
 b) Doppler Institute, Czech Technical University, B\v{r}ehov{\'a} 7,\\
\phantom{e)x}11519 Prague, Czech Republic \\
 c) Department of Mathematics, Faculty of Science, Kanazawa \\
\phantom{e)x}University, Kanazawa 920-1192, Japan \\
 \rm \phantom{e)x}exner@ujf.cas.cz, ichinose@kappa.s.kanazawa-u.ac.jp}
\vspace{8mm}

\noindent {\small We study measure perturbations of the Laplacian
in $L^2(\R^2)$ supported by an infinite curve $\Gamma$ in the
plane which is asymptotically straight in a suitable sense. We
show that if $\Gamma$ is not a straight line, such a ``leaky
quantum wire'' has at least one bound state below the threshold of
the essential spectrum.}
\end{quote}


\section{Introduction}

The aim of the present paper is to elucidate some geometrically
induced spectral properties for the Laplacian in $L^2(\R^2)$
perturbed by a negative multiple of the Dirac measure of an
infinite curve $\Gamma$ in the plane.

This problem has at least two motivations. On the physics side we
note that quantum mechanics of electrons confined to narrow
tubelike regions has attracted a considerable interest, because
such systems represent a natural model for semiconductor ``quantum
wires''. In some examples the region in question is a strip or
tube with hard walls -- see, \eg, \cite{DE} and references therein
-- while other treatments assume even stronger localization to a
curve or a graph -- a rich bibliography to such models can be
found in \cite{KS}. Various interesting spectral effects were
found in such a setting related to the geometry and topology of
the underlying restricted configuration space. One of them, of a
relevance for the present paper, is the existence of
curvature-induced bound states in Dirichlet tubes observed for the
first time more than a decade ago \cite{ES}.

On the other hand, the said models are certainly idealized as far
as the nature of the confinement is concerned. In actual quantum
wires, the electrons are trapped due to interfaces between two
different semiconductor materials which represents a finite
potential jump. Hence if two parts of a quantum wire are close to
each other, a quantum tunneling is possible between them. The
idealization thus makes an important difference, because without
it one expects the spectral properties to be determined by the
{\em global} geometry of the wire. At the same time, it is not
{\em a priori} clear whether effects like the curvature-induced
binding mentioned above will persist if a tunneling is allowed,
because the techniques used to demonstrate them make essential use
of the strict spatial localization.

Here we address the last question in the weak-coupling setting
when the confinement is realized transversally by an attractive
$\delta$ interaction \cite{AGHH}. We will show that if such a
confining interaction is supported by a non-straight curve which
is, however, straight asymptotically in the sense which we make
precise below, the corresponding Hamiltonian has a nontrivial
discrete spectrum. This is our main result expressed by
Theorem~\ref{dsexist}. Moreover, we will show in
Theorem~\ref{approx} that such Hamiltonians can be approximated in
the norm-resolvent sense by a family of Schr\"odinger operators
with regular potentials of the form of a bounded and infinitely
stretched ``ditch''. Consequently, the approximating operators
exhibit bound states too provided the ditch is squeezed enough.

On the other hand, the technique we employ to demonstrate these
results may represent some mathematical interest. It is basically
the Birman-Schwinger formalism in the form extended to
measure-perturbed Laplacians in \cite{BEKS}. In the present case,
however, we deal with the situation where the operator appearing
in the BS-kernel is not compact. Our treatment shows that one can
nevertheless get a useful information, if the operator in question
decomposes into a sum of two parts, of which one is an operator
with a known spectrum and the other is its compact perturbation.


\setcounter{equation}{0}
\section{Generalized Schr\"odinger operators} \label{prelim}

The Hamiltonians we are going to study are generalized
Schr\"odinger operators with a singular interaction supported by a
zero-measure set. Let us first recall several facts about such
operators. They are borrowed from the paper \cite{BEKS} and we
specify them to our present purpose by assuming the configuration
space dimension $d=2$ and the coupling ``strength'' constant on
the interaction support.

Consider a positive Radon measure $m$ on $\R^2$ and a number
$\alpha>0$ such that
\begin{equation} \label{basiccond}
(1+\alpha) \int_{\R^2} |\psi(x)|^2\, \D m(x) \le a \int_{\R^2}
|\nabla\psi(x)|^2\, \D x + b \int_{\R^2} |\psi(x)|^2\, \D x
\end{equation}
holds for all $\psi\in\SSS(\R^2)$ and some $a<1$ and $b$. The map
$I_m$ defined by $I_m\psi=\psi$ on $\SSS(\R^2)$ extends by density
uniquely to
\begin{equation}
I_m:\: W_{1,2}(\R^2) \,\to\, L^2(m):= L^2(\R^2,m) \;;
\end{equation}
for the sake of brevity we employ the same symbol for a continuous
function and the corresponding equivalence classes in both
$L^2(\R^2)$ and $L^2(m)$. The inequality (\ref{basiccond}) extends
to $W_{1,2}(\R^2)$ with $\psi$ replaced by $I_m\psi$ at the \lhs

The operators we are interested in are introduced by means of the
following quadratic form,
\begin{equation} \label{Hamform}
\EE_{-\alpha m}(\psi,\phi) :=  \int_{\R^2}
\overline{\nabla\psi(x)} \nabla\phi(x) \, \D x -\alpha \int_{\R^2}
(I_m\bar\psi)(x) (I_m\phi)(x)\, \D m(x)\,,
\end{equation}
with the domain $W_{1,2}(\R^2)$. It is straightforward to see
\cite{BEKS} that under the condition (\ref{basiccond}) this form
is closed and below bounded, with $C_0^{\infty}(\R^2)$ as a core,
and consequently, it is associated with a unique self-adjoint
operator denoted as $H_{-\alpha m}$. The condition
(\ref{basiccond}) is satisfied, in particular, if the measure $m$
belongs to the generalized Kato class
\begin{equation} \label{Kato}
\lim_{\epsilon\to 0}\: \sup_{x\in\R^2}\, \int_{B(x,\epsilon)} |\ln
|x\!-\!y|| \, \D m(x) = 0\,,
\end{equation}
where $B(x,\epsilon)$ is the ball of radius $\epsilon$ and center
$x$. Moreover, any positive number can be in this case chosen as
$a$.

For operators of the described type the generalized
Birman-Schwinger principle is valid. If $k^2$ belongs to the
resolvent set of $H_{-\alpha m}$ we put $R^k_{-\alpha m} :=
(H_{-\alpha m}-k^2)^{-1}$. The free resolvent $R^k_0$ is defined
for $\im k>0$ as an integral operator with the kernel
\begin{equation} \label{freeG}
G_k(x\!-\!y) = {i\over 4}\, H_0^{(1)} (k|x\!-\!y|)\,.
\end{equation}
Next we need embedding operators associated with $R^k_0$. Let
$\mu, \nu$ be arbitrary positive Radon measures on $\R^2$ with
$\mu(x)= \nu(x) =0$ for any $x\in\R^2$. By $R^k_{\nu,\mu}$ we
denote the integral operator from $L^2(\mu):=L^2(\R^2,\mu)$ to
$L^2(\nu)$ with the kernel $G_k$, \ie
$$ R^k_{\nu,\mu} \phi = G_k \ast \phi\mu $$
holds $\nu$-a.e. for all $\phi\in D(R^k_{\nu,\mu}) \subset
L^2(\mu)$. In our case the two measures will be the $m$ introduced
above and the Lebesgue measure $\D x$ on $\R^2$ in different
combinations. With this notation one can express the generalized
BS principle as follows \cite{BEKS}:
\begin{proposition} \label{BS}
(i) There is a $\kappa_0>0$ such that the operator $I-\alpha
R^{i\kappa}_{m,m}$ on $L^2(m)$ has a bounded inverse for any
$\kappa \ge \kappa_0$. \\ [1mm]
(ii) Let $\im k>0$. Suppose that $I-\alpha R^k_{m,m}$ is
invertible and the operator
$$ R^k := R_0^k + \alpha R^k_{\D x,m} [I-\alpha R^k_{m,m}]^{-1}
R^k_{m,\D x} $$
from $L^2(\R^2)$ to $L^2(\R^2)$ is everywhere defined. Then $k^2$
belongs to $\rho(H_{-\alpha m})$ and $(H_{-\alpha m}-k^2)^{-1}=
R^k$.
\\ [1mm]
(iii) $\:\dim\ker(H_{-\alpha m}-k^2) = \dim\ker(I-\alpha
R^k_{m,m})$ for any $k$ with $\im k>0$.
\end{proposition}


\setcounter{equation}{0}
\section{Formulation of the problem} \label{formul}

After this preliminary we will specify a class of operators which
we discuss in the following, where the measure $m$ will be the
Dirac measure supported by a curve. Suppose that $\tilde\gamma:
\R\to\R^2$ is a {\em continuous, piecewise $C^1$ smooth} function;
its graph is a curve denoted as $\Gamma$. We can define its arc
length,
$$ s[\xi_1,\xi_2]:= \int_{\xi_1}^{\xi_2}
\sqrt{\dot{\tilde\gamma}^2_1 + \dot{\tilde\gamma}^2_2}\: \D\xi\,,
$$
which is the natural parametrization of $\Gamma$: for a fixed
$\xi_1$, $s[\xi_1,\cdot]$ is strictly increasing and piecewise
smooth, so there is a unique inverse function $\xi:\R\to\R$ with
the same properties, and we can define $\gamma:= {\tilde\gamma}
\circ \xi$. In what follows we characterize the curve $\Gamma$
always by the function $\gamma$. Since $\gamma$ maps continuously
into $\R^2$, we have
\begin{equation} \label{geobound}
|\gamma(s)-\gamma(s')| \le |s-s'|
\end{equation}
for any $s,s'\in \R$. In addition, we shall assume:
\begin{description}
 \vspace{-1.2ex}
 \item{\em (a1)} there is $c\in(0,1)$ such that $|\gamma(s)-\gamma(s')|
 \ge c|s-s'|$. In particular, $\Gamma$ has no cusps and
 self-intersections, and its possible asymptotes are not parallel
 to each other.
 \vspace{-1.2ex}
 \item{\em (a2)} $\:\Gamma$ is asymptotically straight in the
 following sense: there are positive $d,\, \mu$, and $\omega\in(0,1)$
 such that the inequality
\begin{equation} \label{asympt}
1-\, {|\gamma(s)-\gamma(s')|\over|s-s'|} \le d \left\lbrack
1+|s+s'|^{2\mu} \right\rbrack^{-1/2}
\end{equation}
 holds true in the sector $S_\omega:= \left\{ (s,s'):\: \omega <
 {s\over s'} < \omega^{-1}\, \right\}$.
\end{description}
The operator we are interested in is a generalized Schr\"odinger
operator with the interaction localized at the curve which can
be formally written as
\begin{equation} \label{Ham}
H_{\alpha, \gamma} = -\Delta -\alpha\delta(x-\Gamma)\,.
\end{equation}
This definition can be given meaning if we identify $H_{\alpha,
\gamma}$ with $H_{-\alpha m}$ of the preceding section, where
$m$ is the Dirac measure on $\Gamma$, or more exactly,
\begin{equation} \label{meas}
m:\; m(M) = \ell_1(M\cap\Gamma)
\end{equation}
for any Borel $M\subset\R^2$, where $\ell_1$ is the
one-dimensional Hausdorff measure; for a piecewise smooth curve it
is given, of course, by the arc length.

One has to make sure, of course, that the measure (\ref{meas})
satisfies the condition (\ref{basiccond}). This follows from
Thm.~4.1 of \cite{BEKS} if $\gamma$ is continuous, piecewise
$C^1$, and satisfies the assumption {\em (a1)}. Consequently, we
may employ Proposition~\ref{BS} for investigation of the
resolvent of $H_{\alpha, \gamma}$.


\setcounter{equation}{0}
\section{Leaky wires as weakly coupled waveguides}
\label{squeeze}

Before proceeding further we want to show that the operators
(\ref{Ham}) can be regarded as weak-coupling approximation to a
class of Schr\"odinger operators. Let $\Gamma$ be again an
infinite planar curve described by the function $\gamma$. Now we
shall make a stronger assumption, namely that $\gamma$ is $C^2$.
Then we can define the (signed) curvature $k(s):= \left(\gamma'_1
\gamma''_2 - \gamma'_2 \gamma''_1 \right)(s)$; we shall assume
that it is bounded, $|k(s)|<c_+$ for some $c_+>0$ and all
$s\in\R$. We employ the conventional symbol believing that the
context will never allow to mix the curvature with the momentum
variable. On the other hand, we will not impose the requirements
{\em (a1), (a2)}. It is sufficient to assume that $\Gamma$ has
neither self-intersections nor ``near-intersections'', \ie, that
there is a $c_->0$ such that $|\gamma(s)\!-\!\gamma(s')|\ge c_-$
for any $s,s'$ with $|s\!-\!s'|\ge c_-$.

Under these assumptions we are able to define in the vicinity of
$\Gamma$ a locally orthogonal system of coordinates: a point is
characterized by the pair $(s,u)$, where $u$ is the (signed)
distance from $\Gamma$ measured along the appropriate normal
$n(s)$ and $s$ is the arc-length coordinate of the point of
$\Gamma$ where the normal is taken. It is easy to see that the
curvilinear coordinates are well defined and unique in the strip
neighbourhood of the curve, $\Sigma_{\epsilon}:= \{ x(s,u):\:
(s,u)\in \Sigma^0_{\epsilon}\}$, where
\begin{equation} \label{curvi}
x(s,u):= \gamma(s) + n(s)u \end{equation}
and $\Sigma^0_{\epsilon}:= \{(s,u):\: s\in\R, |u|<\epsilon\}$ is
the straightened strip, as long as the condition $2\epsilon< c_-$
is valid. If there is no danger of misunderstanding, we shall
write simply $x$ instead of $x(s,u)$.

With these prerequisites we are able to construct the mentioned
family of Schr\"odinger operators. Given $W\in
L^{\infty}((-1,1))$, we define for all $\epsilon< {1\over 2}\,
c_-$ the transversally scaled potential,
\begin{equation} \label{scaledpot}
V_{\epsilon}(x) := \left\lbrace
\begin{array}{ccl} 0 & \quad\dots\quad & x\not\in\Sigma_{\epsilon}
\\ -\,{1\over\epsilon}\, W\left(u\over\epsilon \right) &
\quad\dots\quad & x\in\Sigma_{\epsilon} \end{array} \right.
\end{equation}
and put
\begin{equation} \label{scaledHam}
H_{\epsilon}(W,\gamma):= -\Delta\, +\, V_{\epsilon} \,.
\end{equation}
The operators $H_{\epsilon}(W,\gamma)$ are obviously self-adjoint
on $D(-\Delta)= W_{2,2}(\R^2)$ and the corresponding resolvent can
be expressed in the Birman-Schwinger way,
\begin{eqnarray} \label{BSres}
\lefteqn{ \left(H_{\epsilon}(W,\gamma) \!-\!k^2 \right)^{-1} =
\left(-\Delta \!-\!k^2 \right)^{-1} } \\ && - \left(-\Delta
\!-\!k^2 \right)^{-1} V_{\epsilon}^{1/2} \left\lbrack I +
|V_{\epsilon}|^{1/2} \left(-\Delta \!-\!k^2 \right)^{-1}
V_{\epsilon}^{1/2} \right\rbrack^{-1} |V_{\epsilon}|^{1/2}
\left(-\Delta \!-\!k^2 \right)^{-1} \nonumber
\end{eqnarray}
for any $k^2\in \rho\left(H_{\epsilon}(W,\gamma) \right) \cap
\rho\left(-\Delta \right)$, where we have used the usual
convention, $V_{\epsilon}^{1/2}:= |V_{\epsilon}|^{1/2}
\sgn(V_{\epsilon})$.

Then we have the following approximation result the proof of which
is given in the appendix:
\begin{theorem} \label{approx}
With the stated assumptions, $H_{\epsilon}(W,\Gamma) \to
H_{\alpha, \gamma}$ as $\epsilon\to 0$, where $\alpha =
\int_{-1}^1 W(t)\, \D t$, in the norm-resolvent sense.
\end{theorem}


\setcounter{equation}{0}
\section{Curvature-induced discrete spectrum} \label{curv-bs}

Let us return now to the spectral analysis of the operator
$H_{\alpha,\gamma}$. If $\Gamma$ is a straight line corresponding
to $\gamma_0(s)= as+b$ for some $a,b\in\R^2$ with $|a|=1$, we can
separate variables and show that
\begin{equation} \label{straightspec}
\sigma(H_{\alpha,\gamma_0}) = \left\lbrack -{1\over 4}\alpha^2,
\infty \right)
\end{equation}
is purely absolutely continuous. The aim of the present section is
to show that for a non-straight $\Gamma$ of the class specified in
Sec.~\ref{formul}, $\sigma(H_{\alpha,\gamma})$ has a nonempty
discrete component. Let us start with the essential spectrum.

\begin{proposition} \label{essent}
Let $\alpha>0$ and suppose that $\gamma: \R\to\R^2$ is a
continuous, piecewise $C^1$ function satisfying (a1), (a2); then
$\sigma_{ess}(H_{\alpha,\gamma}) = \left\lbrack -{1\over
4}\alpha^2, \infty \right)$.
\end{proposition}
{\sl Proof:} We shall show in a while that
$\sigma(\RR^{\kappa}_{\alpha,\gamma_0}) =[0,\alpha/2\kappa]$ holds
for $\RR^{\kappa}_{\alpha,\gamma}:= \alpha R^{i\kappa}_{m,m}$
referring to $\gamma=\gamma_0$. In view of Lemma~\ref{HS} below
the same interval is contained in the spectrum of
$\RR^{\kappa}_{\alpha,\gamma}$, and thus by Proposition~\ref{BS}
no point of the interval $\left( -{1\over 4}\alpha^2, 0\right)$
belongs to the resolvent set of the operator $H_{\alpha,\gamma}$.
Consequently, $\sigma_{ess}(H_{\alpha,\gamma}) \supset
\left\lbrack -{1\over 4}\alpha^2, 0\right\rbrack$. By the same
compact-perturbation argument we find that apart of a discrete set
corresponding to eigenvalues of a finite multiplicity, the points
$-\kappa^2$ with $\kappa>\, {1\over 2}\alpha$ belong to
$\rho\left(H_{\alpha,\gamma} \right)$, so the interval
$(-\infty,-{1\over 4}\alpha^2)$ is not contained in the essential
spectrum.

It remains to deal with the positive halfline. First we notice
that for any $R>0$ one can find a disc $B_R\subset \R^2$ of radius
$R$ which does not intersect with $\Gamma$, for otherwise we may
take a family of such discs centered at the points $(3n_1R,0)$ and
$(0,3n_2R)$ with $n_1, n_2\in\Z$, and any curve intersecting with
all of them would violate the assumption {\em (a2)}.

Let $\phi\in C_0^{\infty}([0,2))$ with $\phi(r)\ge 0$ and
$\int_{\R^2} \phi (|x|)^2 dx = 1$. Given $n\in\Z_0$ and
$p,x_n\in\R^2$, we define
$$ \psi_n(x;p,x_n):= \frac{1}{n}
\phi\left(\frac{1}{n}|x-x_n|\right)\, \e^{ipx}\,. $$
The functions $\psi_n$ are normalized and easily seen to provide
for an appropriate sequence $\{x_n\}\subset\R^2$ with $|x_n|
\rightarrow \infty$ a Weyl sequence of the free Hamiltonian $H_0$
corresponding to the point $|p|^2$ of its essential spectrum.
Choosing now the sequence $\{x_n\}$ in such a way that the discs
$B_{2n}(x_n)$ are mutually disjoint and do not intersect with
$\Gamma$, we have $H_{\alpha,\gamma}\psi_n(\cdot;p,x_n)
=H_0\psi_n(\cdot;p,x_n)$. In this way, we have constructed a Weyl
sequence to $H_{\alpha, \gamma}$ for any point of $[0,\infty)$
concluding thus the proof. \quad \QED \vspace{3mm}

Now we can state our main result:
\begin{theorem} \label{dsexist}
Adopt the assumptions of the previous proposition. If the
inequality (\ref{geobound}) is sharp for some $s,s'\in \R$, then
$H_{\alpha,\gamma}$ has at least one isolated eigenvalue below
$-{1\over 4}\alpha^2$.
\end{theorem}
{\sl Proof:} By Proposition~\ref{BS} we look for solutions of the
equation $\RR^{\kappa}_{\alpha,\gamma} \psi=\psi$, where
$\RR^{\kappa}_{\alpha,\gamma}:= \alpha R^{i\kappa}_{m,m}$ is an
integral operator on $L^2(\R)$ with the kernel
$$ \RR^{\kappa}_{\alpha,\gamma}(s,s') = {\alpha\over 2\pi}\, K_0
\left( \kappa|\gamma(s)\!-\! \gamma(s')| \right) \;; $$
here $K_0$ is the Macdonald function; recall that $K_0(z)= {\pi
i\over 2} H_0^{(1)}(iz)$. The idea is to compare this
operator with $\RR^{\kappa}_{\alpha,\gamma_0}$ having the kernel
in which $|\gamma(s)\!-\! \gamma(s')|$ is replaced by $|s\!-\!
s'|$.

The Fourier transformation takes $K_0(\kappa x)$ to $(\pi/2)^{1/2}
(p^2+\!\kappa^2)^{-1/2}$. The well known relation $f(-i\nabla)\psi
= (2\pi)^{-1/2} (\FF^{-1}f) \ast \psi$ then shows that
$\RR^{\kappa}_{\alpha,\gamma_0}$ is unitarily equivalent to the
operator of multiplication by ${1\over 2} \alpha
(p^2+\!\kappa^2)^{-1/2}$ on $L^2(\R)$. Consequently, it is
absolutely continuous and its spectrum is $[0,\alpha/2\kappa]$ in
correspondence with (\ref{straightspec}).

We can obtain the spectrum of $H_{\alpha,\gamma_0}$ directly, of
course, as pointed out above. Now we we want to know how the
spectrum of $\RR^{\kappa}_{\alpha,\gamma_0}$ changes under the
perturbation $\DD_{\kappa}:= \RR^{\kappa}_{\alpha,\gamma} -
\RR^{\kappa}_{\alpha,\gamma_0}$. Notice that
\begin{equation} \label{Dkern}
\DD_{\kappa}(s,s') := {\alpha\over 2\pi}\, \bigg( K_0 \left(
\kappa|\gamma(s)\!-\! \gamma(s')| \right) - K_0 \left(
\kappa|s\!-\!s'| \right) \bigg) \ge 0 \end{equation}
holds for the kernel of $\DD_{\kappa}$ in view of
(\ref{geobound}) and the monotonicity of $K_0$.
\begin{lemma} \label{vari}
$\;\sup \sigma\left( \RR^{\kappa}_{\alpha,\gamma} \right) >
{\alpha\over 2\kappa}\;$ if $\,\Gamma$ is not straight.
\end{lemma}
{\sl Proof:} It is sufficient to find a real-valued $\psi\in
\SSS(\R)$ such that
$$ \left(\psi, \RR^{\kappa}_{\alpha,\gamma} \psi \right) -
{\alpha\over 2\kappa} \|\psi\|^2 > 0\,, $$
which is equivalent to
\begin{eqnarray*}
&& {2\kappa\over\alpha} \int_{\R^2} \DD_{\kappa}(s,s')\,
\psi(s)\psi(s')\, \D s\, \D s' \\ && +\, {\kappa\over\pi}
\int_{\R^2} K_0 \left( \kappa|s\!-\!s'| \right) \psi(s)\psi(s')\,
\D s\, \D s' - \int_{\R} \psi(s)^2 \D s
> 0\,. \end{eqnarray*}
Using the above observation together with the Parseval relation
we can rewrite the last two terms at the \rhs as
$$ \int_{\R} {\kappa\over\sqrt{p^2\!+\kappa^2}}\,
|\hat\psi(p)|^2\, \D p - \int_{\R} |\hat\psi(p)|^2\, \D p\,. $$
Choosing
$$ \psi(s) = \sqrt[4]{2\lambda^2\over\pi}\; e^{-\lambda^2
s^2}\,, $$
we find by a direct computation that two terms equal
$$ -\,{1\over \sqrt{2\pi}}\: \int_{\R} \left( 1\!-
{\kappa\over\sqrt{u^2 \lambda^2\! +\kappa^2}} \right) e^{-u^2/2}\,
\D u = -\,{1\over \sqrt{2\pi}}\: {\lambda^2\over 2\kappa^2}\:
\int_{\R} u^2\, e^{-u^2/2}\, \D u + \OO(\lambda^3)\,. $$
On the other hand, the inequality in (\ref{Dkern}) is sharp in
an open subset of $\R^2$ if $\Gamma$ is not straight, so the
first term is
$$ \sqrt{2\over\pi}\:\lambda\, \int_{\R^2} {2\kappa\over\alpha}\,
\DD_{\kappa}(s,s')\, e^{-\lambda^2(s^2\!+s'^2)} \, \D s\,\D s' \ge
c\lambda $$
for some $c>0$ as $\lambda\to 0+$. Hence the above $\psi$ is the
sought trial function for $\lambda$ small enough. \quad \QED
\vspace{2mm}

Next we shall show that perturbation (\ref{Dkern}) is compact
under the assumption {\em (a2)}, and thus it can change only the
discrete spectrum of $\RR^{\kappa}_{\alpha,\gamma}$.
\begin{lemma} \label{HS}
$\;\DD_{\kappa}\;$ is Hilbert-Schmidt if $\mu>\,{1\over 2}\,$.
\end{lemma}
{\sl Proof:} For the sake of brevity, we denote
$$ \varrho \equiv \varrho(s,s') :=
\kappa|\gamma(s)\!-\!\gamma(s')|\,, \quad \sigma \equiv
\sigma(s,s') := \kappa|s\!-\!s'|\,. $$
To estimate $K_0(\varrho) -K_0(\sigma)$ we use convexity of $K_0$
together with the relation $K'_0(z)= -K_1(z)$,
\begin{equation} \label{Dest}
K_1(\sigma) (\sigma\!-\!\varrho) \le K_0(\varrho)
-K_0(\sigma) \le \varrho K_1(\varrho) {\sigma\!-\!\varrho \over
\varrho} \,. \end{equation}
Hence the kernel of $\DD_{\kappa}$ is bounded, because $\varrho
\mapsto \varrho K_1(\varrho)$ is bounded in $(0,\infty)$ and the
inequality $c\sigma\le \varrho\le\sigma$ yields
\begin{equation} \label{geo}
0\le {\sigma\!-\!\varrho \over \varrho} \le {1\!-\!c \over
c}\,. \end{equation}
Moreover, there is $c_1>0$ such that
\begin{equation} \label{Kest}
\varrho K_1(\varrho) \le c_1\, e^{-\varrho/2} \le c_1\,
e^{-c\sigma/2}\,,
\end{equation}
and by {\em (a2)} we have
\begin{equation} \label{geo2}
{\sigma\!-\!\varrho \over \varrho} \le {\sigma\!-\!\varrho
\over c\sigma} \le {d\over c}\, \left\lbrack 1+|s+s'|^{2\mu}
\right\rbrack^{-1/2} \end{equation}
in the sector $S_\omega$. Putting together the inequalities
(\ref{Dest})--(\ref{geo2}) we can estimate the Hilbert-Schmidt
norm of the operator in question:
\begin{eqnarray} \label{HSnorm}
\lefteqn{ \left(2\kappa\over\alpha \right)^2 \int_{\R^2}
\DD_{\kappa}(s,s')^2\, \D s\, \D s' \le  \left(1\!-\!c \over
c\right)^2\, c_1^2 \int_{\R^2 \setminus S_{\omega}}
e^{-c\kappa|s-s'|} \, \D s\, \D s'}
\\ && +\, \left(c_1 d\over c\right)^2\: \int_{S_{\omega}}
{e^{-c\kappa|s-s'|} \over 1+|s+s'|^{2\mu} }\, \D s\, \D s'
\nonumber \\ && \le \left(2c_1 {1\!-\!c \over c}\right)^2
{1+\omega \over 1-\omega} \int_0^{\infty} u\, e^{-\sqrt{2} c\kappa
u} \, \D u + \left(c_1 d\over c\right)^2\! \int_{\R^2}
{e^{-c\kappa|s-s'|} \over 1+|s+s'|^{2\mu} }\, \D s\, \D s',
\nonumber
\end{eqnarray}
which is finite for $\mu>\, {1\over 2}\,$. \quad \QED
\vspace{2mm}

Finally, we need the following continuity result.
\begin{lemma} \label{cont}
With the above stated assumptions, the function $\;\kappa\mapsto
\RR^{\kappa}_{\alpha,\gamma}\,$ is operator-norm continuous and
$\,\RR^{\kappa}_{\alpha,\gamma}\to 0 $ as $\kappa\to\infty$.
\end{lemma}
{\sl Proof:} Using the above established equivalence between
$\RR^{\kappa}_{\alpha,\gamma_0}$ and the multiplication by
${1\over2}\, \alpha[p^2\!+\kappa^2]^{-1/2}$ we easily check the
claim for the ``free'' operator, so it is sufficient to show
that the perturbation $\DD_{\kappa}$ has the same properties.
The inequality
$$ \left|\left( \DD_{\kappa}-\DD_{\kappa'} \right)(s,s')
\right|^2 \le 2\left\lbrack \DD_{\kappa}(s,s')^2 +\DD_{\kappa'}
(s,s')^2 \right\rbrack \le 4\DD_{\kappa_0}(s,s')^2  $$
valid for any $\kappa_0 \le \min(\kappa,\kappa')$ allows us to
use the dominated convergence by which
$$ \| \DD_{\kappa}-\DD_{\kappa'} \|_\mathrm{HS} \to 0 \qquad
\mathrm{as} \qquad \kappa'\to\kappa\,.  $$
Finally, the estimate (\ref{HSnorm}) shows at the same time that
$$ \| \DD_{\kappa} \|_\mathrm{HS} \to 0 \qquad \mathrm{as}
\qquad \kappa\to\infty\,,  $$
which concludes the proof. \quad \QED \vspace{2mm}

\noindent {\sl Proof of Theorem~\ref{dsexist}, continued:} By
Lemma~\ref{vari} $\;\sup \sigma\left( \RR^{\kappa}_{\alpha,\gamma}
\right) > {\alpha\over 2\kappa}$ holds whenever $\Gamma$ is not
straight. On the other hand, the essential spectrum of
$\RR^{\kappa}_{\alpha,\gamma_0}$ is by Lemma~\ref{HS} preserved
under the geometric perturbation, so
$\RR^{\kappa}_{\alpha,\gamma}$ has in $\left({\alpha\over
2\kappa}, \infty \right)$ just isolated eigenvalues; in
combination with the previous result we infer that at least one
such eigenvalue $\lambda_{\alpha,\gamma}(\kappa)$ of
$\RR^{\kappa}_{\alpha,\gamma}$ exists for any $\kappa>0$. Finally,
by Lemma~\ref{cont} the function $\lambda_{\alpha,\gamma}(\cdot)$
is continuous and $\lim_{\kappa\to\infty}
\lambda_{\alpha,\gamma}(\kappa)=0$. Hence there is a point
$\kappa_0 >\, {1\over 2}\alpha$ such that
$\lambda_{\alpha,\gamma}(\kappa_0)=1$, and therefore, recalling
that $\RR^{\kappa}_{\alpha,\gamma}= \alpha R^{i\kappa}_{m,m}$, we
infer by Proposition~\ref{BS} that $\;-\kappa_0^2$ is an
eigenvalue of the operator $H_{\alpha,\gamma}$. \quad \QED

\begin{remark} \label{compar}
{\rm One asks naturally how strong is the asymptotic restriction
imposed by {\em (a2)}. To answer this question, suppose that
$\gamma$ is $C^2$ smooth. The $\Gamma$ can be described --
uniquely up to Euclidean transformations of the plane -- by its
signed curvature $k(s)$. Using the standard expression of $\gamma$
in terms of $k$ we can estimate
\begin{eqnarray*}
\lefteqn{|\gamma(s) - \gamma(s')| = \Bigg\lbrack \left(
\int_{s'}^s \cos \left( \int_{s'}^{s_1} k(s_2)\, \D s_2 \right) \D
s_1 \right)^2 }
\\ && +\, \left( \int_{s'}^s \sin \left( \int_{s'}^{s_1} k(s_2)\,
\D s_2 \right) \D s_1 \right)^2 \Bigg\rbrack^{1/2} \\ && \ge
\int_{s'}^s \cos \left( \int_{s'}^{s_1} k(s_2)\, \D s_2 \right) \D
s_1 \ge \int_{s'}^s \left\lbrack 1- {1\over 2} \left(
\int_{s'}^{s_1} k(s_2)\, \D s_2 \right)^2 \right\rbrack \D s_1\,,
\end{eqnarray*}
where we have assumed $s>s'$ without loss of generality; hence
$$ 1-\, {|\gamma(s)-\gamma(s')|\over|s-s'|} \le {1\over
2|s\!-\!s'|} \int_{s'}^s \left( \int_{s'}^{s_1} k(s_2)\, \D s_2
\right)^2 \D s_1\,. $$
Suppose that $|k(s)|\le c_2|s|^{-\beta}$ for some $\beta>0$, then
the \rhs of the last inequality can be estimated by
$$ {1\over 2|s\!-\!s'|}\: {c^2_2\over |s|^{2\beta}}\:  \int_{s'}^s
(s_1\!-\!s')^2 \D s_1 \,\le\, {c^2_2\over |s'|^{2\beta}}\:
{|s\!-\!s'|^2\over 6} \,\le\, {c^2_2 s^2\over 6|s'|^{2\beta}}
\,\le\, {c^2_2 \over 6\omega^2}\: |s'|^{2-2\beta}\,. $$
Consequently, {\em (a2)} with $\mu >{1\over 2}$ holds for $\beta>
{5\over 4}$. This is a slightly stronger restriction than for
curved Dirichlet strips \cite{DE} where $\beta>1$ is sufficient. }
\end{remark}


\setcounter{section}{1} \setcounter{equation}{0}
\renewcommand{\theequation}{\Alph{section}.\arabic{equation}}
\renewcommand{\theclaim}{\Alph{section}.\arabic{equation}}
\section*{Appendix}

To prove Theorem~\ref{approx} we have to show that (\ref{BSres})
approximates the resolvent of the formal operator (\ref{Ham})
which we have identified with $H_{-\alpha m}$. We will write the
resolvents in question in a way similar to that used for the
analogous purpose in \cite[Sec.~I.3]{AGHH}. The first term at the
\rhs of (\ref{BSres}) is $\epsilon$-independent and subtracts in
the difference. The action of the second one on a vector $\psi\in
L^2(\R^2)$ can be written as
\begin{eqnarray}
\lefteqn{ - \int\!\!\int\!\!\int_{\R^2} G_k( x\!-\! x')
V_{\epsilon}^{1/2}( x') \left\lbrack I + |V_{\epsilon}|^{1/2}
R^k_0 V_{\epsilon}^{1/2} \right\rbrack^{-1} ( x', x'')
|V_{\epsilon}|^{1/2}( x'') } \nonumber \\ && \phantom{AAAA} \times
G_k( x''\!-\! x''')\, \psi( x''')\, \D x'\, \D x''\, \D x'''
\nonumber \\ &&
= \int\!\!\int_{\Sigma}\int_{\R^2} G_k\left( x\!-\! x(s',u')
\right)\, {1\over\epsilon}\, W^{1/2}\left({u'\over\epsilon}
\right) \nonumber \\ && \phantom{AAAA} \times \epsilon
\left\lbrack I + |V_{\epsilon}|^{1/2} R^k_0 V_{\epsilon}^{1/2}
\right\rbrack^{-1} (s',u';s'',u'')\, {1\over\epsilon} \left|
W\left( {u''\over\epsilon} \right) \right|^{1/2}  \nonumber \\ &&
\phantom{AAAA} \times G_k\left( x'''\!-\! x(s'',u'') \right)\,
(1+u'k(s')) (1+u''k(s'')) \nonumber
\\ && \phantom{AAAA} \times \psi( x''')\, \D s'\, \D u'\, \D
s''\, \D u''\, \D x''' \label{2ndres}
\end{eqnarray}
where $x(s,u)$ is given by (\ref{curvi}) and $(1+uk(s))$ is the
Jacobian of the transformation between the Cartesian and
curvilinear coordinates. Changing the integration variables to
$t':= u'/\epsilon$ and $t'':= u''/\epsilon$ we can rewrite the
last expression as
\begin{eqnarray*}
\lefteqn{\int\!\!\int_{\Sigma}\int_{\R^2} G_k\left( x\!-\!
\gamma(s') \!-\! n(s') \epsilon t'\right)\, W^{1/2}\left( t'
\right) }
\\ && \phantom{AA} \times \epsilon \left\lbrack I +
|V_{\epsilon}|^{1/2} R^k_0 V_{\epsilon}^{1/2} \right\rbrack^{-1}
(s',\epsilon t';s'',\epsilon t'')\, \left| W\left( t'' \right)
\right|^{1/2} \\ && \phantom{AA} \times G_k\left( x'''\!-\!
\gamma(s'') \!-\! n(s'') \epsilon t''\right)\, (1+ \epsilon
t'k(s')) (1+ \epsilon t''k(s''))
\\ && \phantom{AA} \times \psi( x''')\, \D s'\, \D u'\, \D
s''\, \D u''\, \D x'''\,.
\end{eqnarray*}
If $\||V_{\epsilon}|^{1/2} R^k_0 V_{\epsilon}^{1/2}\| < 1$, the
inverse can be written as a geometric series with the
integral-operator kernel
\begin{eqnarray*}
\lefteqn{ \epsilon \left\lbrack I + |V_{\epsilon}|^{1/2} R^k_0
V_{\epsilon}^{1/2} \right\rbrack^{-1} (s',\epsilon t';s'',\epsilon
t'') } \\ && = \delta(s'\!-\!s'')\, \delta(t'\!-\!t'') -
|W(s',t')|^{1/2} G_k(s', \epsilon t'; s'', \epsilon t'')
W(s'',t'')^{1/2} + \dots
\end{eqnarray*}
Consequently, the operator given by (\ref{2ndres}) can be written
as the product $B_{\epsilon} (I\!-\!C_{\epsilon})^{-1} \tilde
B_{\epsilon}$ of operators mapping $L^2(\R^2) \to L^2(\Sigma^0_1)
\to L^2(\Sigma^0_1) \to L^2(\R^2) $, with the following kernels
\begin{eqnarray*}
B_{\epsilon}( x; s',t') &\!:=\!& G_k\left( x\!-\! x(s',\epsilon
t')\right)\, (1+ \epsilon t'k(s')) W\left(t' \right)^{1/2} \,, \\
\tilde B_{\epsilon}(s,t; x') &\!:=\!& |W\left(t \right)|^{1/2} (1+
\epsilon t k(s))\, G_k\left( x'\!-\! x(s,\epsilon t)\right),
\\ C_{\epsilon}(s,t;s',t') &\!:=\!& |W\left(t
\right)|^{1/2} G_k(x(s,\epsilon t) \!-\!x( s', \epsilon t'))
W\left(t' \right)^{1/2} \,.
\end{eqnarray*}
We have $\|C_{\epsilon}\| \le \|W\|_{\infty} \|P_1 R_0^k P_1\| \le
\|W\|_{\infty} |k|^{-2}$ for $k=i\kappa$ with $\kappa>0$, where
$P_1$ is the projection onto $L^2(\Sigma^0_1) \subset L^2(\R^2) $,
hence $\|C_{\epsilon}\|\le \mathrm{const} <1$ holds for $\kappa$
large enough uniformly w.r.t. $\epsilon$, and the operator in
question equals
\begin{equation} \label{epsres}
B_{\epsilon} (I\!-\!C_{\epsilon})^{-1} \tilde B_{\epsilon} =
\sum_{j=0}^{\infty} B_{\epsilon} C_{\epsilon}^j \tilde
B_{\epsilon}\,.
\end{equation}

Let us now turn to the resolvent of $H_{\alpha,\gamma}$. Since the
operator $I\!-\!\alpha R^k_{m,m}$ is by Proposition~\ref{BS}
boundedly invertible with for $k=i\kappa$ with $\kappa$ large
enough, we can again write its second terms as a geometric series.
Furthermore, $\alpha= (W^{1/2}, |W|^{1/2})$ by assumption, so we
have
\begin{eqnarray}
\alpha \lefteqn{ R^k_{\D x,m} \sum_{j=0}^{\infty} \left( \alpha
R^k_{m,m} \right)^j R^k_{m,\D x} = R^k_{\D x,m} (W^{1/2},
|W|^{1/2}) R^k_{m,\D x} } \nonumber \\ && + R^k_{\D x,m} (W^{1/2},
|W|^{1/2}) R^k_{m,m} (W^{1/2}, |W|^{1/2}) R^k_{m,\D x} + \dots
\nonumber \\ && = \sum_{j=0}^{\infty} B C^j \tilde B\,,
\label{zerores}
\end{eqnarray}
where $B,C,\tilde B$ are operators between the same spaces as
their indexed counterparts above given by their integral kernels:
\begin{eqnarray*}
B( x; s',t') &\!:=\!& G_k\left( x\!-\! \gamma(s')\right)\,
W\left(t' \right)^{1/2} \,, \\ \tilde B(s,t; x') &\!:=\!&
|W\left(t \right)|^{1/2}\, G_k\left( x'\!-\! \gamma(s)\right),
\\ C(s,t;s',t') &\!:=\!& |W\left(t
\right)|^{1/2} G_k(\gamma(s) \!-\!\gamma( s')) W\left(t'
\right)^{1/2} \,.
\end{eqnarray*}
Let us stress that while these operators depend on $W$, the
expression (\ref{zerores}) contains just the integral of the
approximating potential, which is why the limit does not depend on
a particular shape of $W$. The operator norm of the difference
between (\ref{epsres}) and (\ref{zerores}) can be estimated by
means of the telescopic trick,
\begin{eqnarray*}
\lefteqn{ \left\| B_{\epsilon} (I\!-\!C_{\epsilon})^{-1} \tilde
B_{\epsilon} - B(I\!-\!C)^{-1} \tilde B \right\| \le
\sum_{n=0}^{\infty} \Big\lbrace \| B_{\epsilon}\!-\!B\|
\|C_{\epsilon}\|^n \|\tilde B_{\epsilon}\| } \\ && + \|B\|
\sum_{\ell=0}^{n-1} \|C\|^{\ell}  \| C_{\epsilon}\!-\!C\|
\|C_{\epsilon}\|^{n-\ell-1} \|\tilde B_{\epsilon}\| + \|B\|
\|C\|^n \|\tilde B_{\epsilon}\!-\!\tilde B\| \Big\rbrace\,,
\end{eqnarray*}
where the second term at the \rhs is conventionally put to zero
if $n=0$. As above, we have $\|R^k_0\|\le |k|^{-2}$ for
$-ik=\kappa>0$, with $\|W^{1/2}\| \le \|W\|^{1/2}_{\infty}$ and
$|1\!+\!\epsilon t k(s)| \le 1+ \epsilon \|k\|_{\infty} < 1+
\|k\|_{\infty}$, hence for $k^2$ large enough negative there is
a positive $c_3<1$ such that
$$ \max\{ \|B\|, \| B_{\epsilon}\|, \|C\|, \| C_{\epsilon}\|,
\|\tilde B\|, \|\tilde B_{\epsilon}\| \} \le c_3 $$
holds for any $\epsilon\in(0,1)$. Consequently, the norm in
question is estimated by
$$ \left\lbrace \| B_{\epsilon}\!-\! B\| + \|\tilde
B_{\epsilon}\!-\!\tilde B\| \right\rbrace \sum_n c_3^{n+1} +
\|C_{\epsilon}\!-\!C\| \sum_n n\,c_3^{n+1}\,, $$
so it is sufficient to investigate the three norms involved here.
Consider the first one which we can estimate as follows,
$$\| B_{\epsilon}\!-\! B\| \le  \|W\|^{1/2}_{\infty} \left\lbrace
(1\!+\! \|k\|_{\infty}) \left\| R^k_{\Sigma,\epsilon} \!-\!
R^k_{\Sigma,0} \right\| + \epsilon \|k\|_{\infty} \left\|
R^k_{\Sigma,0} \right\| \right\rbrace\,, $$
where $R^k_{\Sigma,\epsilon}, R^k_{\Sigma,0}$ are the resolvent
factors in this expression, \ie, integral operators
$L^2(\Sigma^0_1) \to L^2(\R^2)$ with kernels $G_k\left( x\!-\!
x(s',\epsilon t')\right)$ and $G_k\left( x\!-\!
\gamma(s')\right)$, respectively. To show that $R^k_{\Sigma,
\epsilon} \to R^k_{\Sigma,0}$ in the operator-norm topology, let
us rewrite the kernel of the difference by means of the mean value
theorem,
\begin{eqnarray*}
&& G_k\left( x\!-\! x(s',\epsilon t')\right)- G_k\left( x\!-\!
\gamma(s')\right)
\\ && \hspace{-6mm} = {1\over 2\pi} \bigg\lbrack K_0\left(
\kappa| x\!-\! x(s',\epsilon t')|\right)- K_0\left(\kappa| x\!-\!
\gamma(s')|\right) \bigg\rbrack
\\ && \hspace{-6mm} = -\, {\epsilon t'\over 2\pi}\, \int_0^1  K_1\left(
\kappa| x\!-\! \gamma(s') \!-\! n(s')\epsilon t'
\vartheta|\right)\, \kappa \left( {d\over d\vartheta}\,
\mathrm{dist} (x,\gamma(s') \!+\! n(s')\epsilon t'\vartheta)
\right)\, \D\vartheta\,.
\end{eqnarray*}
Since the last factor does not exceed
one in modulus, we have
\begin{equation}
\left| \left( R^k_{\Sigma,\epsilon} \!-\! R^k_{\Sigma,0}\right)
(x,x(s', \epsilon t') \right| \le {\epsilon\kappa |t'|\over
2\pi}\, \int_0^1 K_1\left( \kappa| x\!-\! \gamma(s') \!-\!
n(s')\epsilon t'\vartheta |\right)\, \D\vartheta\,.
\end{equation}
This makes it possible to estimate the quantity
\begin{eqnarray*}
\lefteqn{h_{\infty}:= \sup_{x\in\R^2} \int_{\R} \D s' \int_{-1}^1
\D t' \left| \left( R^k_{\Sigma,\epsilon} \!-\!
R^k_{\Sigma,0}\right) (x,x(s', \epsilon t') \right| }
\\ && \le {\epsilon\kappa\over 2\pi}\, \sup_{x\in\R^2} \int_{\Sigma^0_1}
K_1\left( \kappa| x\!-\! x(\sigma') )|\right) \D\sigma'
\\ && \le {\epsilon\kappa\over 2\pi}\, \sup_{x\in\R^2} \int_{\R^2}
K_1\left( \kappa| x\!-\! x') )|\right) \D x' = {\epsilon\kappa\over
2\pi}\, \left\| K_1(\kappa|\cdot|)\right\|_{L^1(\R^2)}\,,
\end{eqnarray*}
where the \rhs is finite, because the function
$K_1(\kappa|\cdot|)$ decays exponentially at large distances and
has the integrable singularity $|\cdot|^{-1}$ at the origin. In
the same way we find
$$ h_1:= \sup_{x'\in\Sigma_1} \int_{\R^2} \left| \left(
R^k_{\Sigma,\epsilon} \!-\! R^k_{\Sigma,0}\right) (x,x') \right|
\D x \le {\epsilon\kappa\over 2\pi}\, \left\|
K_1(\kappa|\cdot|)\right\|_{L^1(\R^2)}\,. $$
The norm under consideration can be the estimated by the
corresponding Schur-Holmgren bound -- see, \eg,
\cite[Ex.~III.3.2]{Ka} -- as
$$\left\|R^k_{\Sigma, \epsilon} - R^k_{\Sigma,0} \right\| \le
\left( h_1 h_{\infty}\right)^{1/2} \le {\epsilon\kappa\over 2\pi}\,
\left\| K_1(\kappa|\cdot|)\right\|_{L^1(\R^2)}\,, $$
so it tends to zero as $\epsilon\to 0$. Analogous estimates are
valid for $\|\tilde B_{\epsilon}\!-\!\tilde B\|$ and
$\|C_{\epsilon}\!-\!C\|$ which concludes the proof. \\[2mm]
{\bf Remark:} With our goal in mind we examined the situation when
the approximating potential depends on the transverse variable
only. If we replace it by $W\in L^{\infty}(\Sigma^0_1)$, the
analogous argument shows that corresponding family
(\ref{scaledHam}) converges in the norm-resolvent sense to the
operator $-\Delta+ \alpha(s) \delta(x\!-\!\gamma(s))$ with
$\alpha(s):= \int_{-1}^1 W(s,u)\, \D u$, which is properly defined
by a quadratic form similar to (\ref{Hamform}) -- see \cite{BEKS}.


\subsection*{Acknowledgments}

P.E. and T.I. are respectively grateful for the hospitality
extended to them at the University of Kanazawa and at the Nuclear
Physics Institute, AS CR, where parts of this work were done. The
research has been partially supported by GAAS and Czech Ministry
of Education under the contracts 1048801 and ME170.

\end{document}